# Spin Vector Control for Heisenberg-Inspired Probabilistic Computing

Yuanqiu Tan ‡,[1,2] Rahul Tripathi ‡, [1,2] Saleh Bunaiyan,[4,5] Ryan Wagner,[3] Neil Dilley,[2] Kerem Camsari,[4] Joerg Appenzeller,[1,2] and Zhihong Chen[*1,2]

[1]*Electrical and Computer Engineering, Purdue University, West Lafayette, Indiana, USA*

[2]*Birck Nanotechnology Center, Purdue University, West Lafayette, Indiana, USA*

[3]*Mechanical Engineering, Purdue University, West Lafayette, Indiana, USA*

[4]*Electrical and Computer Engineering, University of California, Santa Barbara, California, USA*

[5]*Engineering Department, King Fahd University of Petroleum & Minerals (KFUPM), Dhahran 31261, Saudi Arabia*

**To whom correspondence should be addressed.*

*E-mail: zhchen@purdue.edu*

## Abstract

Probabilistic bits (p-bits) have emerged as a cornerstone of probabilistic computing, enabling energy-efficient hardware implementation for probabilistic inference and combinatorial optimization. A critical challenge in advancing this field beyond binary p-bits lies in realizing and manipulating vector spin information, essential for mapping complex energy-based models such as the Heisenberg Hamiltonian. Here, we demonstrate a spintronic platform capable of real-space vector summation by using dual ferromagnetic spin injections into a monolayer graphene channel. By electrically tuning the spin polarization through independently controlled injection currents, we achieve continuous control over the magnitude and direction of the resulting spin accumulation vector. Experimental observations, supported by theoretical vector summation models and spin-circuit simulations, reveal coherent vector interactions and angular tunability of the spin state. This approach enables direct implementation of vector-based spin logic and lays the groundwork for mapping classical Heisenberg models using stochastic low-barrier magnets. Our results establish a scalable pathway for realizing probabilistic spin circuits based on two-dimensional materials, offering new opportunities for low-power, non-Boolean computing architectures.

## Introduction

Conventional computers rely on stable, deterministic bits and charge-based logic systems to perform operations; however, they face critical bottlenecks as Moore's Law slows and transistor scaling reaches its physical limits. Concurrently, the explosive growth in compute-intensive fields such as machine learning, and big-data analytics is driving demand for energy efficient computing power[1]. These disciplines demand rapid improvements to handle the computational requirements of large-scale implementations. In response to these challenges, recent decades have seen a surge in research focused on innovative solutions such as the use of two-dimensional materials for transistors and interconnects[2–4], monolithic 3D integration[5], novel material discovery[6], and in-memory computing[7].

One particularly promising direction is energy-based models (EBMs), which can naturally be mapped to device physics. EBMs with binary variables (Ising models) have been widely explored in spintronic hardware, with reported orders-of-magnitude improvements in area and energy relative to conventional digital implementations[8]. Utilizing spintronics probabilistic bits (p-bits) with two states governed by Ising model, prior works have successfully demonstrated probabilistic inference and combinatorial optimization[9–12]. Extending beyond the binary p-bits,  p-bits governed by classical Heisenberg model with continuous states could further expand computation capabilities, including for training modern Hopfield networks[13,14]. These devices can leverage the vector nature of spin currents, with potential advantages in area, speed, and power efficiency[15]. It is worth mentioning that a digital implementation of the simpler Ising spin model would consume around two orders of magnitude more energy than the spintronic realization[8]; therefore, a digital implementation of the classical Heisenberg model is expected to require even higher energy consumption.

Notably, recent implementations of spin majority gates have demonstrated Boolean logic by summing the scalar projections of spin polarizations from multiple ferromagnetic inputs in a graphene spin channel[16]. However, these approaches only capture a limited aspect of spin behavior, overlooking the full vectorial nature of spin currents that could be harnessed for richer logic and computational operations.

The Heisenberg interaction, which describes the exchange interaction between localized spins, can indirectly influence angle-dependent spin transport through spin mixing experiments. Yet, to observe Heisenberg-like interactions via spin transport, it is critical to understand how these interactions modulate spin dynamics and accumulation in real space. We probe these effects directly using the non-local spin valve (NLSV) devices presented in our work.

A crucial but unresolved challenge in spin-based computing is the controlled vector summation of multiple spin polarizations within a single nonmagnetic channel[17], a functionality that lacks a direct analog in conventional charge-based electronics. The classical notion that spin accumulations from multiple injectors simply superimpose as scalar quantities has been foundational in many experimental interpretations[12,18–20]. Yet, theoretical models inspired by Heisenberg-like spin interactions and stochastic Landau–Lifshitz–Gilbert (LLG) dynamics suggest that vectorial spin-spin coupling, mediated via spin current interactions, can realize continuous-variable computation schemes. These models posit that collective dynamics of low-barrier magnets (LBMs)[21], exchanging spin currents through a spin-neutral channel[22], can minimize a classical Heisenberg Hamiltonian and implement vector-valued summations of magnetic states, effectively serving as hardware solvers for energy-based problems[14,23].

In this work, we report the first comprehensive experimental realization and analysis of a spintronic device that directly performs real-space addition of two spin vectors, not merely scalar projections within a nonmagnetic medium. Our platform enables continuous control over the orientation and magnitude of the resulting spin accumulation vector via electrical tuning of the injection currents and magnetization orientations. Unlike previous implementations of spin majority gates or scalar summation using orthogonal or collinear magnetic configurations, our device performs true vector addition of spin states and probes the resultant accumulation using a unique geometrically engineered injector. This experimental demonstration validates longstanding theoretical predictions about vector-based spin summation and provides a foundational building block for next-generation analog and probabilistic spin logic architectures.

**Statistical Analysis of Non-Local Spin Valve Signals and Electrical Properties in Monolayer Graphene Devices with Ferromagnetic Py Contacts**

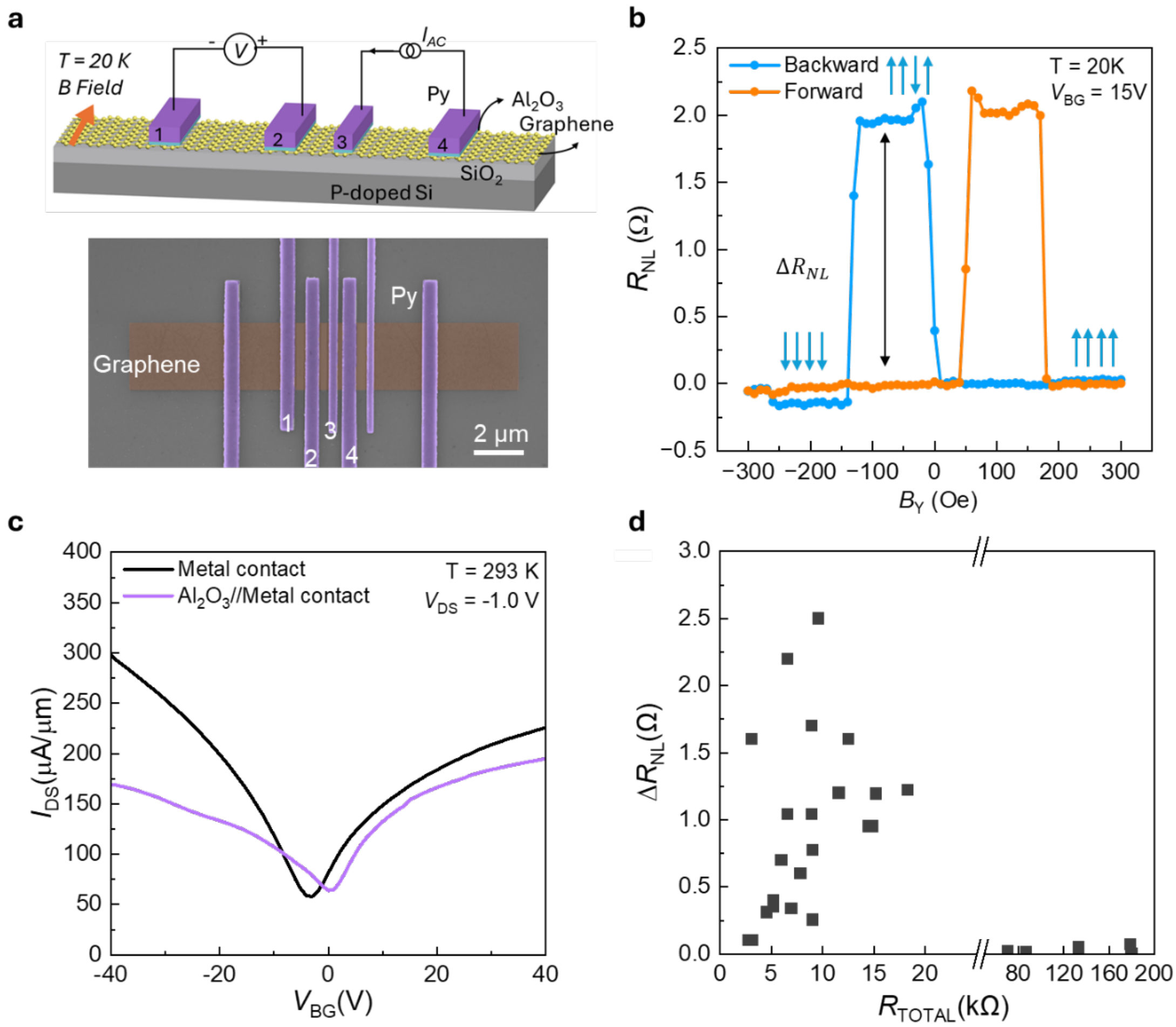

**Fig. 1. Statistics of non-local spin valve signals and electrical characteristics with monolayer graphene. a.** Schematic and scanning electron microscope (SEM) of a lateral non-local spin valve device, showing ferromagnetic permalloy (Py) spin injector and detector electrodes 2 and 3 on top of a patterned graphene channel. An $Al_2O_3$ tunneling barrier is used between the ferromagnets and the graphene channel. **b.** Non-local spin valve signal of a representative device with diffusion length of 500 nm, measured at 20 K with back-gate of 15 V. **c.** Transfer characteristics of representative monolayer graphene devices with/without $Al_2O_3$ tunneling barrier, at $V_{DS}$ = -1.0 V. **d.** Statistical correlation between the non-local spin signal and the total local resistance between injector and detector electrodes, including the graphene channel resistance and two contact resistances.

Efficient spin injection and long diffusion length in graphene spin valves are essential for achieving efficient correlations in LBM arrays without dipolar interference and with low-power injection [18,24]. To accomplish electrical spin injection into a monolayer graphene channel, we fabricated lateral NLSV devices, as shown in Fig. 1a. Permalloy (Py) was used as the ferromagnetic (FM) metal contact for all four electrodes. Considering the conductance mismatch between graphene channel and the FM contacts, a thin Al layer was deposited and subsequently naturally oxidized in ambient conditions to form a tunneling barrier, thereby improving the spin injection efficiency[25]. The detailed device fabrication process is described in the Methods section.

The spin transport properties of all devices were measured at 20 K using standard lock-in techniques. Depending on the direction of the bias magnetic field, the majority spins are injected from FM electrode 3 to 4 via a bias current of 1 μA (Fig. 1a), leading to minority spin accumulation beneath electrode 3 at the Py-graphene interface. These minority spins diffuse leftward through the graphene channel and are detected at FM detector electrode 2, whose magnetization direction determines the spin-dependent voltage signal[26].

The FM contacts were designed with their easy axis aligned along the long edge and present different coercive fields ($H_C$) determined by the width as governed by shape anisotropy. A back-gate voltage was applied to improve the carrier density in graphene (details in the Methods section). When an in-plane magnetic field was swept along the easy axis of the magnets, the magnetization of the ferromagnetic electrodes switched between parallel (P) and antiparallel (AP) configurations[27]. The resulting potential difference between ferromagnetic detect electrodes 2 and 1, is leading to characteristic step-like changes in the non-local resistance ($R_{NL}$) (Fig. 1b). Fig. S1 presents the robust reciprocal relations of the graphene lateral non-local spin signals. Through repeated measurements, alternating the charge spin injection between electrode 3/4 and electrode 1/2 consistently yields identical $R_{NL}$ values, indicating that the same spin signal is achieved regardless of whether electrodes 2 and 3 serve as the injector or the detector. This consistency also demonstrates the device's reliability across multiple measurements.

Fig. 1c compares the transfer characteristics of monolayer graphene with different source/drain contacts. All electrical characterizations were conducted at room temperature under vacuum with a pressure of ~$10^{-5}$ Torr. Devices with the optimized $Al_2O_3$ tunnel barrier show increased contact resistance at a given gate voltage compared to metal-only contacts. This improved interface resistance enhances spin injection by better matching the spin impedance of the FM contacts to the graphene channel and improving interfacial spin polarization.

To evaluate spin transport performance across devices, we statistically analyzed the correlation between total electrical resistance ($R_{Total}$) and non-local resistance ($\Delta R_{NL}$). The total resistance is extracted from transfer characteristics between injector and detector electrodes as follows:

$$R_{Total} = 2R_{Contact} + R_{Graphene\ channel} = \frac{V_{DS}}{I_{DS}}$$

As shown in Fig. 1d, the devices exhibit optimal $\Delta R_{NL}$ for effective spin injection and transport when the $R_{Total}$ is in the 2-20 kΩ range. This rapid electrical correlation approach enables the efficient screening of devices with sufficient spin injections, thereby accelerating the design and optimization of large-scale graphene spin logic integration.

## Spin Vector Control Using Dual Injection

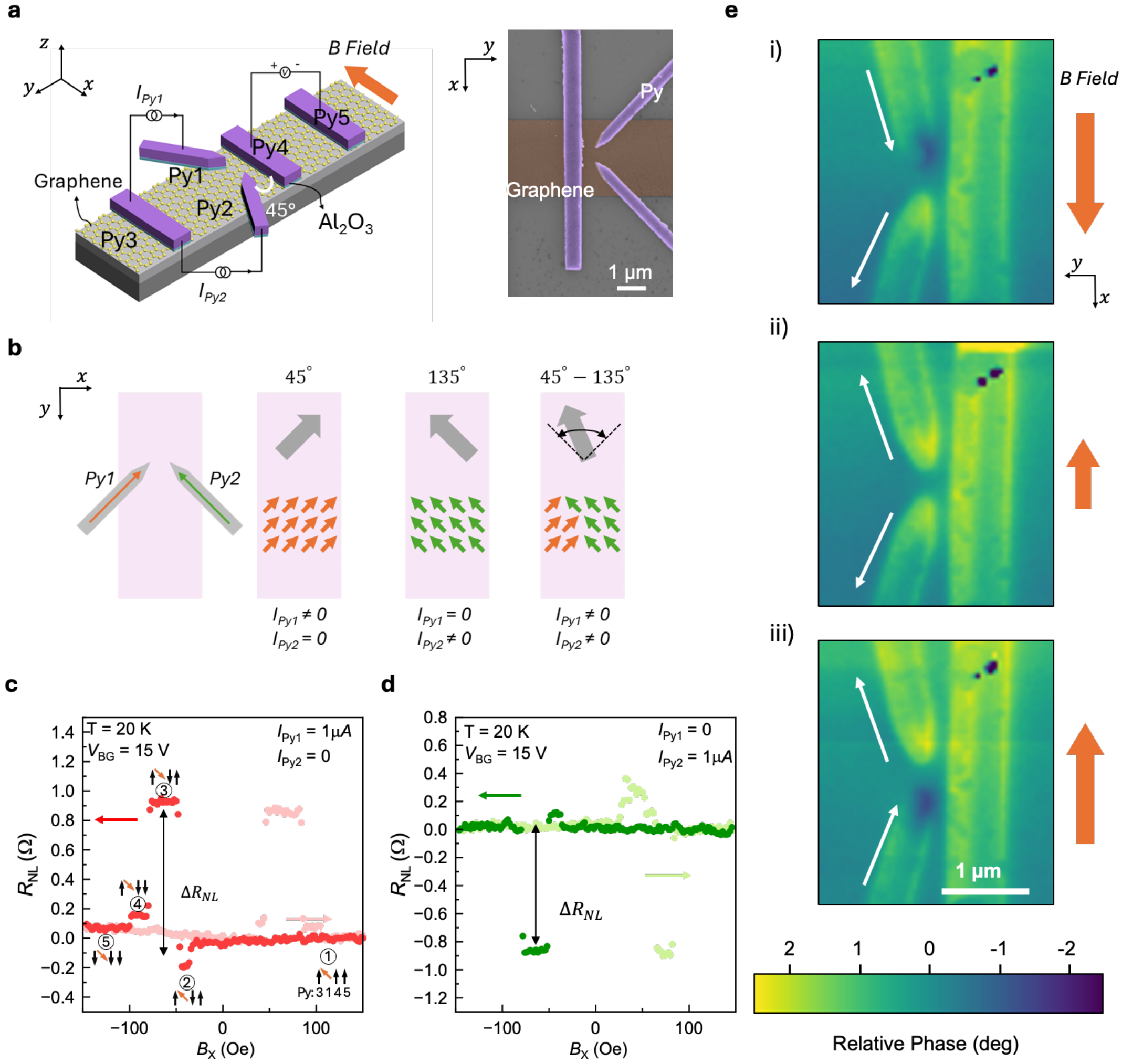

**Fig. 2. Spin vector injection in a graphene spin valve with tilted dual ferromagnetic electrodes. a.** Schematic and SEM of a lateral monolayer graphene NLSV device, with Py FM electrodes and $Al_2O_3$ tunneling barrier. The dual injection electrodes are tilted to 45°/135° with respect to the detector electrode, Py4. **b.** Schematic illustrations of spin vector addition, where the final spin polarization angle is determined by the injection currents from the dual injectors (Py1 and Py2). **c**. Spin signal with only current injection through Py1, at T = 20 K and $V_{BG}$ = 15 V. The $\Delta R_{NL}$ associated with the magnetization reversal of Py1 corresponds to the transition between the ② and ③ resistance states. The magnetization directions of the FM electrodes at each state are indicated for a backward magnetic-field sweep from positive to negative values. See Supplementary Fig. S2 for the corresponding electrochemical potential profiles. **d**. Nonlocal spin signal with only current injection through Py2. The $\Delta R_{NL}$ associated with the switching of Py2 is labeled. **e**. MFM phase images of the dual injection electrodes under different in-plane magnetic fields. The white arrows indicate the relative magnetization direction of Py1 and Py2.

The directions and magnitudes of the applied in-plane magnetic fields are illustrated on the side.

Demonstrating vector operations in spin transport is a critical step toward scalable integration of spin-based probabilistic computation models. We achieve spin vector control by simultaneously injecting spin currents from two ferromagnetic injector electrodes. Fig. 2a illustrates a schematic and scanning electron microscope (SEM) image of a graphene lateral NLSV device with tilted FM injectors, Py1 and Py2. Each injector is designed to be 300 nm wide and 40 nm thick, with edges patterned into needle shapes to mitigate the impact of the demagnetizing field in the remanent state. The remaining three FM electrodes, each 500 nm wide, have their easy axes aligned along the $x$ direction. The two electrodes for injection, Py1 and Py2, are oriented at 45° and 135° relative to the easy axis of the detection electrode Py4. The graphene channel for spin diffusion between the tips of the injectors to the detector electrode is ~500 nm. The design concept is inspired by the work of Kimura *et al.*[17], who demonstrated spin orientation modulation using a Cu diffusion channel.

Although the tilt angles of the injectors are already fixed, the resultant spin summation depends on the relative injector currents $I_1$ and $I_2$. A uniform electrical control of the polarization direction of the injected out-of-equilibrium spin distribution is achieved by the geometric dual injection design. By varying the electrical currents, the direction of resulting spin polarization in the graphene channel can be continuously tuned between $\widehat{M}_1$ and $\widehat{M}_2$. The resultant spin polarization generated by dual injection can be written as:

$$\widehat{m}(I_1, I_2) = \beta_1 \widehat{M}_1 + \beta_2 \widehat{M}_2,$$

where the coefficients are determined by the relative magnitudes of the two injection currents:

$$\beta_{1,2} = \frac{I_{1,2}}{I_1 + I_2}$$

As depicted in Fig. 2b, the angle of the resulting spin vector from the two injectors, $\emptyset\left(\widehat{m}(I_1, I_2)\right)$, can be continuously tuned between 45° and 135°. When the current is applied to only one injector, the spin vector is determined solely by the magnetization direction of that electrode: the spin vector aligns with $\widehat{M}_1$ (45°) when $I_2 = 0$, and with $\widehat{M}_2$ (135°) when $I_1 = 0$. When both injectors are active, the spin vector diffusing through the graphene channel and detected by Py4 corresponds to the vector summation of the spin contributions from the two injectors, with its final direction set by the relative current magnitudes. As an example in the illustration, when the injection from Py1 is smaller than that from Py2, the spin vector lies between 90° to 135°.

Before progressing to a more detailed discussion of spin mixing, it is essential to demonstrate individual spin injection from each injector. Fig. 2c and 2d display the non-local spin valve signals obtained when current is injected between Py1 and Py3 and between Py2 and Py3. During the measurement, the in-plane magnetic field was swept between -150 Oe and 150 Oe. The non-local resistance ($R_{NL}$) measured at Py4 reflects the difference in chemical potentials when the spin accumulation is parallel and antiparallel to the magnetization of Py4. For the backward sweep in Fig. 2c, the $R_{NL}$ transition from the second state to the third, as labeled, occurs when the chemical potential changes due to the magnetization reversal of Py1(detailed illustration in Supplementary Fig. S2). The remaining $R_{NL}$ changes are governed by the magnetization switching of the three parallel electrodes. A similar spin transport behavior is observed when spin is injected from Py2. Similar $H_C$ related to magnetization reversal of each FM can be observed when current is injected separately from Py1 and Py2. Py1 and Py2 generate polarized spin vectors at 45° and 135°, respectively, leading to opposite signs of $\Delta R_{NL}$ when their spin vectors are projected onto the easy axis of the detector Py4. Additionally, when the injection currents are equal ($I_1 = I_2$) and the magnetic field is weak, the resulting signals from Py1 and Py2 are symmetric but inverted, reflecting the 90° angular separation between the magnetization directions of the two injectors. Importantly, the magnitude of the spin signal, $\Delta R_{NL}$, from both injectors is similar, indicating the integrity of spin injection and uniform diffusion in the channel.

Since an external magnetic field is applied during the spin valve measurements, it is important to track how the magnetic moments of the two tilted spin-injection electrodes evolve under different field conditions. To do so, we performed magnetic force microscopy (MFM) measurements while gradually increasing the in-plane external magnetic field applied along the direction parallel to the detector electrode Py4. To initialize the magnetization, the field was first set to +200 Oe, aligning all ferromagnetic electrodes along their easy axes. As the field was gradually swept to –200 Oe, MFM was performed at each step to capture both the topography and phase contrast. In Fig. 2e, the two injectors (Py1 and Py2) exhibit distinct coercive fields due to geometric differences. The phase image (ii) shows the sequential magnetization reversals of the electrodes as the field is swept. Notably, at sufficiently small fields, the magnetization directions of Py1 and Py2 do not necessarily align with the applied field direction.

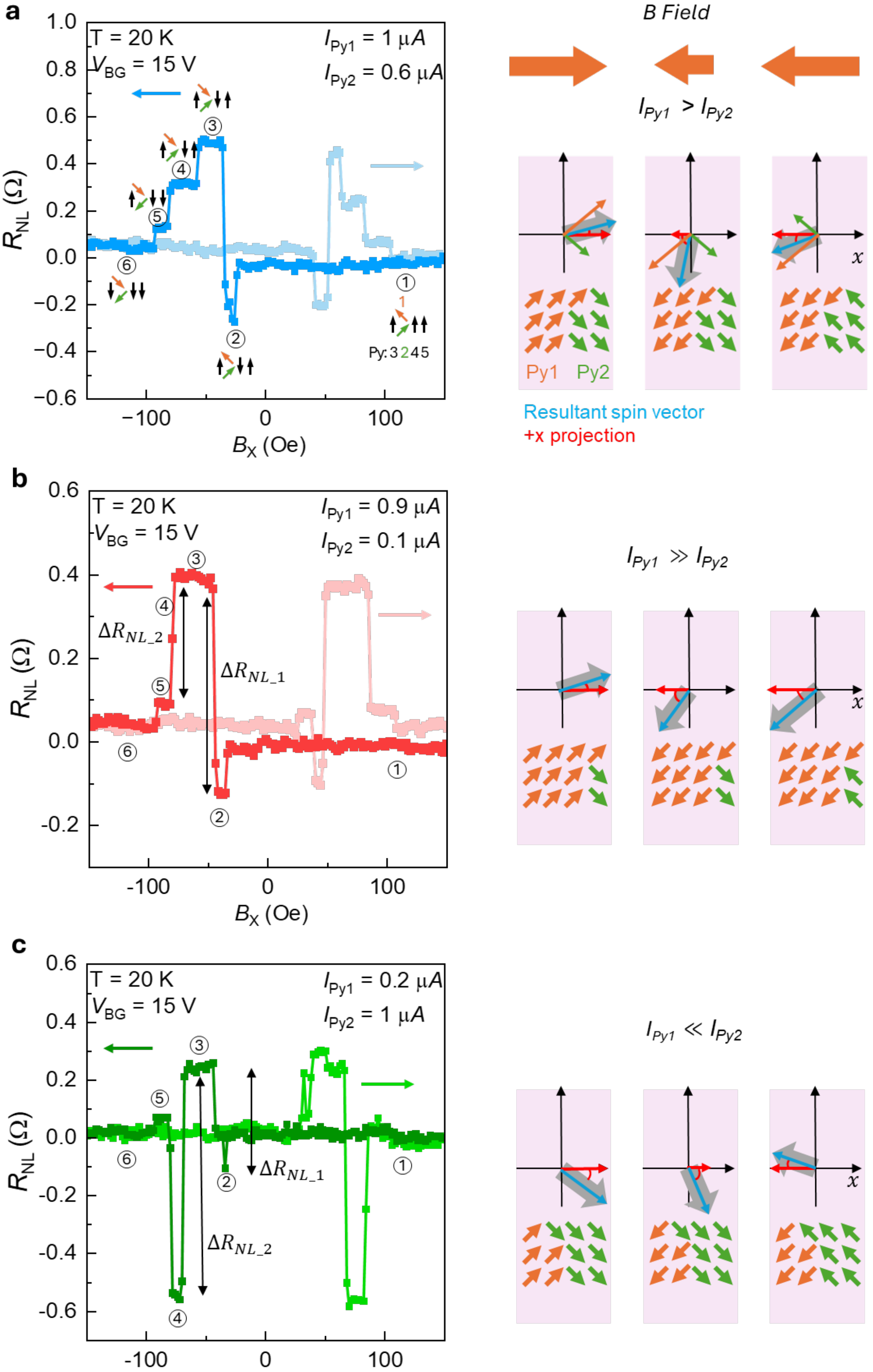
a
T = 20 K
V_BG = 15 V
I_Py1 = 1 μA
I_Py2 = 0.6 μA
R_NL (Ω)
B_X (Oe)
Py: 3 2 4 5
B Field
I_Py1 > I_Py2
Py1
Py2
Resultant spin vector
+x projection
b
T = 20 K
V_BG = 15 V
I_Py1 = 0.9 μA
I_Py2 = 0.1 μA
ΔR_NL_2
ΔR_NL_1
I_Py1 ≫ I_Py2
c
T = 20 K
V_BG = 15 V
I_Py1 = 0.2 μA
I_Py2 = 1 μA
I_Py1 ≪ I_Py2

**Fig. 3. Spin signal modulation via non-collinear dual injection a.** Non-local spin signal under comparable dual injection conditions: $I_{Py1}$ = 1 μA, and $I_{Py2}$ = 0.6 μA. The magnetization directions of the FM electrodes at each of the six resistance states are indicated for a backward magnetic-field sweep from positive to negative values. The accompanying vector illustrations depict the evolution of the resultant spin vector (blue vector) and projection (red vector) at three different magnetic fields, corresponding to the switching sequence observed in the MFM maps of Fig. 2e, with a slightly stronger spin injection from Py1. **b.** Non-local spin signal with Py1 dominant injection: $I_{Py1}$ = 0.9 μA, and $I_{Py2}$ = 0.1 μA. The vector illustration shows a reduced angular rotation and a smaller change in the spin vector projection across the middle and final switching events. **c.** Non-local spin signal with $I_{Py1}$ = 0.2 μA, and $I_{Py2}$ = 1 μA. The vector illustration indicates that the resultant spin vector rotates into the fourth quadrant during the intermediate switching stage. All spin transport measurements were performed at T = 20 K and $V_{BG}$ = 15 V.

Fig. 3a-c presents representative non-local spin valve signals under varying dual spin injection conditions. The spin-vector summation schematic illustrations on the right depict the spin orientations under different magnetic fields and injector currents. In Fig. 3a, with a relatively balanced injection from Py1 and Py2 ($I_{Py1}$ = 1 μA, and $I_{Py2}$ = 0.6 μA), six distinct resistance states are observed, with each resistance change corresponding to the magnetization reversal of one of the five FM electrodes. Consistent with the single injection case, the transitions between the second and third resistance states and between the third and fourth states arise from the magnetization switching of Py1 and Py2, while the other transitions are due to the magnetization switching of the three parallel FM electrodes. Initially, under a large positive in-plane magnetic field, both Py1 and Py2 are magnetized along the $+x$ direction, generating a spin accumulation vector in the first quadrant. As the field is swept to negative, Py1 switches first (around −30 Oe), rotating the resultant spin vector into the third quadrant. Upon further increase of the negative field, Py2 switches (around −60 Oe), leading to a larger magnetization in the third quadrant (see Supplementary for details). These sequential transitions are consistent with the switching behavior observed in the MFM phase imaging (see Fig. 2e). A positive $\Delta R_{NL}$ is observed, reflecting the slightly stronger contribution from Py1. The spin signal exhibits symmetric behavior during the reverse sweep, indicating reproducible and coherent switching dynamics.

In Fig. 3b, where the injection is dominated by Py1 ($I_{Py1}$ = 0.9 μA, and $I_{Py2}$ = 0.1 μA), only a single resistance state change associated with switching of Py1 is observed. The imbalance in injection current results in an initial spin accumulation strongly aligned with the magnetization of Py1. When Py1 switches, the resultant spin vector undergoes a sharp rotation, however, the contribution from Py2 is insufficient to produce a noticeable resistance change. Consequently, the angular rotation of the spin vector between the intermediate and final magnetization states becomes negligible in this limit, effectively collapsing resistance between the third, fourth and fifth states into one. The resultant spin signal shows behavior similar to the single Py1 injection observed in Fig. 2c. Different current injection configurations into Py1 and Py2 generate different Oersted fields, which in turn lead to variations in the $H_C$ of Py1 and Py2 observed in the measurements.

In contrast, Fig. 3c shows the spin transport characteristic when the injection is dominated by Py2 ($I_{Py1}$ = 0.2 μA, and $I_{Py2}$ = 1 μA), showing distinct five resistance transitions. In this case, the resultant spin vector rotates to the fourth quadrant after Py1 switches. After Py2 subsequently switches, the vector rotates into the second quadrant. The resistance changes associated with Py1 and Py2 show negative $\Delta R_{NL}$, consistent with single-injection signal from Py2.

Overall, the variation in $\Delta R_{NL}$ amplitudes and polarities across different injection configurations reflects the tunable spin accumulation achieved via vector summation. The angular dependence and magnitude of these spin vectors, extracted from the labeled $\Delta R_{NL}$ values, provide the foundation for further analysis in the next section.

## Theoretical Vector Model and Angular Dependence

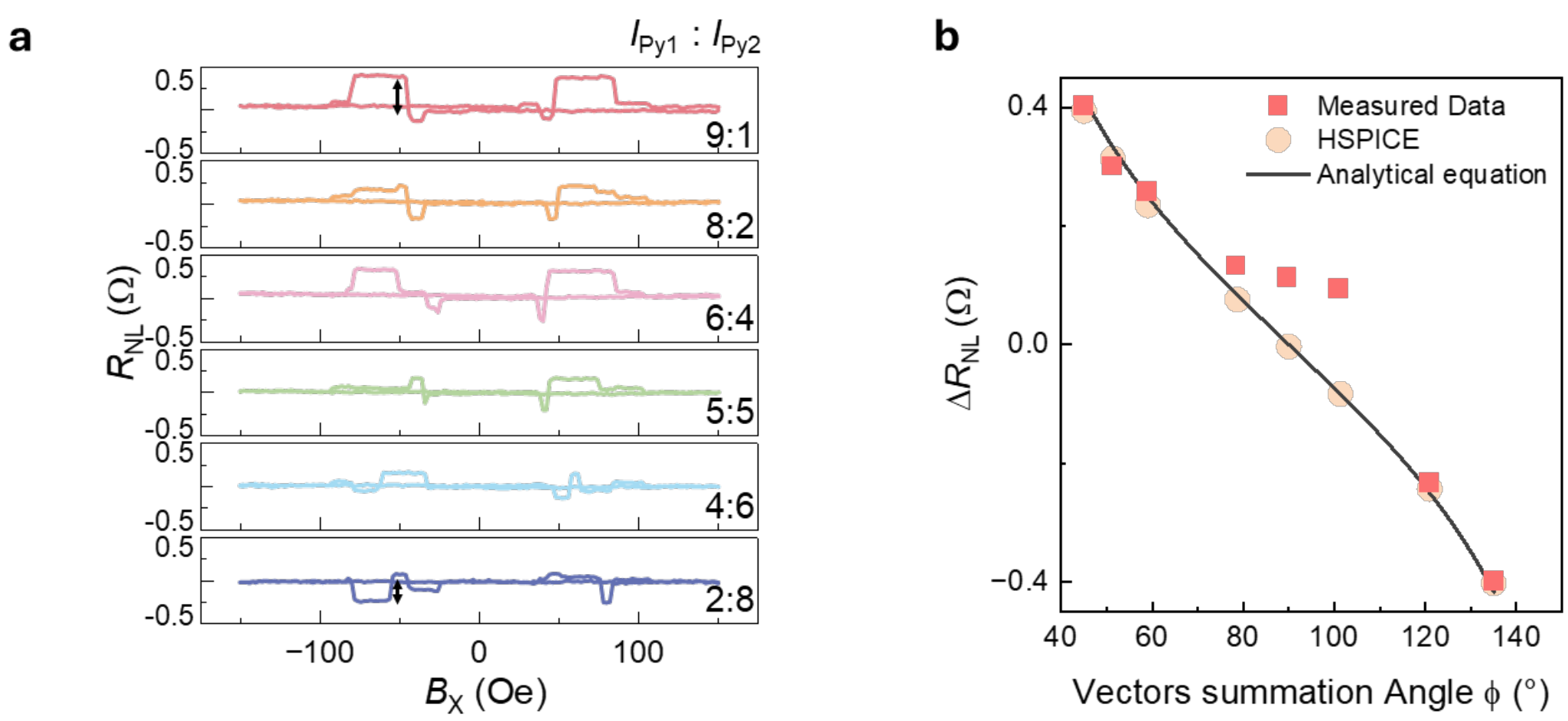

**Fig. 4. Non-collinear detection and signal dependence on injection ratio. a.** Non-local spin valve signals measured by varying the relative current contributions from Py1 and Py2, with the total injection current held constant. The resultant spin accumulation vector evolves continuously as a function of injection ratio (values labeled). **b**, Corresponding non-local spin signal $\Delta R_{NL}$ (squares) plotted as a function of the spin vector angle $\emptyset$, extracted from the data in panel **a**, illustrating the angular tunability of the accumulated spin under fixed total current. The solid line shows analytical calculations. The orange dots show the HSPICE-simulated non-local spin signal responses $\Delta R_{NL}$ (dots) as a function of the angle $\emptyset$. All spin transport measurements were conducted with T = 20 K and $V_{BG}$ = 15 V.

We readily see that the non-local resistance ($\Delta R_{NL}$) reflects the spin accumulation associated with the magnetization rotation of the resultant spin vector injected by Py1 and Py2, as well as the switching of the FM detect electrodes. Owing to the tilted dual injection geometry, the resulting net spin accumulation is generally not collinear with the detector magnetization. Consequently, the dependence of $\Delta R_{NL}$ on the injection ratio can be captured by a vector summation model relating non-local resistance to the spin vector summation angle ($\emptyset$) as [17]:

$$\Delta R_{NL} = \frac{V_P - V_{AP}}{I} = \frac{\Delta V(\emptyset)}{I_s(\emptyset)/p}$$

where

$$I_s(\emptyset) = I(P_1\beta_1\widehat{M}_1 + P_2\beta_2\widehat{M}_2)$$

$I_s(\emptyset)$ is the magnitude of the spin current and $P_{1,2}$ is the interface spin polarization of the injectors. And we assume uniform spin diffusion in the graphene channel.

In order to map the full vector space enabled by dual spin, the non-local spin valve signals are measured by varying the proportion of $I_1$ and $I_2$ under a fixed total injection current ($I_1 + I_2$), as shown in Fig. 4a. Maintaining a constant total current effectively removes the contribution from overall current magnitude, allowing the measured signal evolution to predominantly reflect the control of the spin vector direction through ${}^{I_1}/_{I_2}$ ratio. $\Delta R_{NL}$ is positive when the spin injection from Py1 exceeds that from Py2, i.e. ${}^{I_1}/_{I_2} > 1$. As the proportion of spin injections from Py2 increases, $\Delta R_{NL}$ gradually decreases and switches sign, eventually reaching absolute values comparable to the single-injection limits at $I_2 = 0$ or $I_1 = 0$. By tuning the bias current through the dual injectors, we can precisely control both the magnitude and the angle of the total spin vector injected into the graphene channel.

The consistent evolution of the non-local spin valve signals confirms the symmetric and continuous control of the spin vectors by dual injection. The angle $\emptyset$ of the net spin vector is given by:

$$\emptyset = \tan^{-1}\left(\frac{P_1\beta_1 \sin\theta_1 + P_2\beta_2 \sin\theta_2}{P_1\beta_1 \cos\theta_1 + P_2\beta_2 \cos\theta_2}\right)$$

Where $\theta_{1,2}$ denote the injector Py1 and Py2's magnetization orientations relative to the detector axis.

Based on the vector summation model, we derive detailed approximate analytical expressions for our dual injection configuration, revealing the non-collinear dependence of the spin signals on current injections[28,29] (see details in supplementary).

Given by the vector summation of two spin vectors oriented at 45° and 135°, $\vec{\varphi}(\emptyset) = \overrightarrow{\varphi_1}(\theta_1) + \overrightarrow{\varphi_2}(\theta_2)$, is used to visualize the evolution of the total spin polarization direction. The experimentally measured non-local resistance $R_{NL}$ at each injection configuration is extracted (Fig. 4b). When the total injected current is held constant, increasing the proportion of the 135°-oriented component $\overrightarrow{\varphi_2}(\theta_2)$, relative to the 45°-oriented component $\overrightarrow{\varphi_1}(\theta_1)$ causes the resultant spin vector $\vec{\varphi}(\emptyset)$ to rotate smoothly from 45° toward 135°, following a $\cos^{-1}$ dependence[30,31]. The $\cos^{-1}$ function reflects the projection of the spin vector along the x-axis, which corresponds to the easy axis of ferromagnetic detector electrode.

Experimentally extracted $R_{NL}$ values present excellent agreement with analytical calculations, where the same spin polarizations for dual-injectors and detector are assumed. Small deviations are observed near $\emptyset \sim 90°$, which emerge from the vectorial interplay between the two spin injectors oriented at 45° and 135°. This flat region arises when the x-component of the net spin accumulation projected along the detector axis is minimized due to near-complete cancellation between the oppositely directed contributions from the two injectors. Physically, this corresponds to a regime where the resultant spin polarization vector becomes nearly orthogonal to the detector magnetization, thereby suppressing the measurable signal. Such cancellation condition is achieved when the weighted cosine projections of the spin currents from both injectors are approximately equal in magnitude but opposite in sign. This effect represents a geometric suppression of the spin signal without requiring spin dephasing or relaxation, and it underscores the critical role of angular alignment and polarization asymmetry in vector spin logic architectures. In the angular range from 75° to 105°, both $R_{NL}$ and its modulation are intrinsically small, leading to an increased measurement uncertainty and a reduced signal-to-noise ratio.

The analytical model is further validated by spin-circuit simulations using HSPICE (see Methods and Supplementary). The close correspondence across measurement, calculation, and simulation establishes a robust framework for describing dual-injection spin dynamics in vector domain.

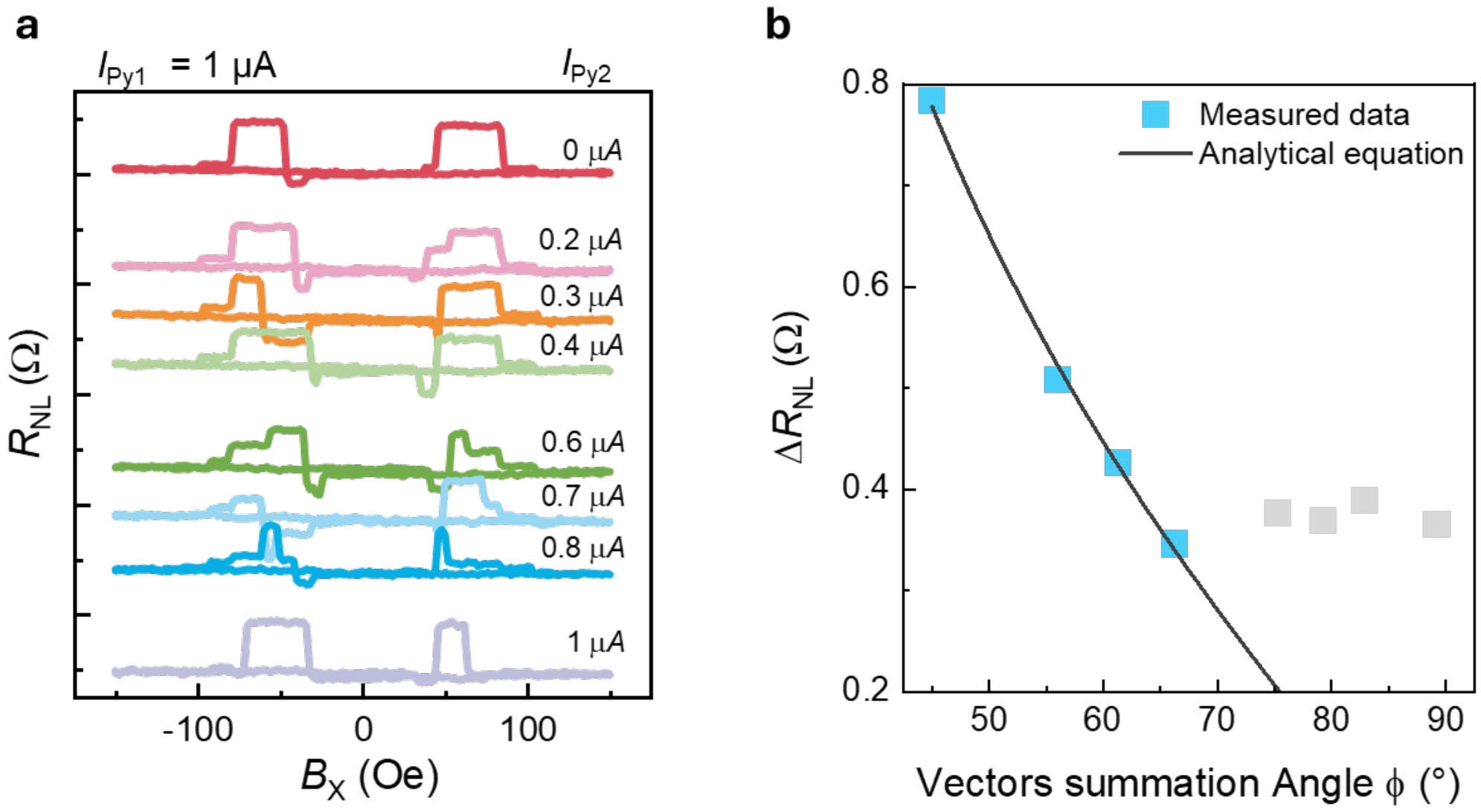

**Fig. 5. Non-collinear detection and signal dependence on injection ratio and total magnitude. a.** Non-local spin valve signals obtained by varying the injection current into Py2 ($I_{Py2}$) while keeping the Py1 current fixed ($I_{Py1}$ = 1 μA). This configuration enables independent tuning of both the magnitude and orientation of the resultant spin vector (currents labeled). **b**, Corresponding non-local spin signal $\Delta R_{NL}$ (squares) plotted as a function of the spin vector angle ∅, extracted from the data in panel **a**. The solid line presents the calculated non-local spin signals from analytical equations. Grey points were excluded from the comparison; see Supplementary Fig. S3 for a detailed discussion. All spin transport measurements were conducted with T = 20 K and $V_{BG}$ = 15 V.

To further generalize the dual-injection behavior, we extended our analysis by varying both the relative proportion and total magnitude of the injection currents. In Fig. 5a, the current injection in Py2, $I_{Py2}$, is modulated while maintaining a constant current injection in Py1, $I_{Py1}$. Starting from single injection in Py1, the resultant $\Delta R_{NL}$ is positive; as the current injection in Py2 increases, $\Delta R_{NL}$ decreases. The resulting changes in $\Delta R_{NL}$ reflect both the current amplitudes and vector summation.

In Fig. 5b, a distinct plateau (grey squares) is experimentally observed in $R_{NL}$ as a function of the effective spin injection angle. This plateau originates from the same suppression mechanism observed in Fig. 4b, where the signal approaches zero and is ultimately limited by the experimental measurement sensitivity. As the 135° component dominates more, the resultant spin vector approaches the predicted orientation, and the measured $\Delta R_{NL}$ shows excellent agreement with the analytical model (Fig. 5b). Moreover, since the total injected current increases in the measurements presented in Fig. 5, the resulting curvature in $\Delta R_{NL}$ is noticeably smoother than in Fig.4, where the total injection current is held constant. This reflects an enhanced sensitivity to both the magnitude and direction angle of spin accumulation, capturing simultaneous amplitude modulation and angular rotation of the resultant vector.

Fig. S3 shows the non-local spin valve signals in the complementary case in which the current injected from Py1 is varied while current injected from Py2 is fixed. Although the absolute values do not match quantitatively, the measurement results agree qualitatively with the analytical calculations and HSPICE simulations. The observed behavior in all cases exemplifies the core principle of vector addition that governs spin transport dynamics.

Furthermore, we extend our dual-injection approach to an alternative injection configuration to examine spin vector summation under different angular conditions (15° /165°) in Fig. S6. This alternative configuration enables a distinct mode of spin vector summation and further demonstrates the robustness and tunability of our dual-injection method for manipulating spin orientations over a broader vector space. The consistent behavior observed across different injector geometries reinforces the validity of the vector summation model in describing non-collinear spin transport in monolayer graphene and confirms that spin transport retains coherence under controlled injection conditions.

## Conclusion

In this work, we have experimentally demonstrated controlled vector spin injection and real-space summation in monolayer graphene spin valves through a dual-injector architecture. By engineering the injector geometry and precisely tuning the injection current ratios, we achieved continuous modulation of the spin polarization direction within the graphene channel. The measured non-local spin signals are in excellent agreement with theoretical predictions based on vector summation models and HSPICE spin-circuit simulations, confirming that the resultant spin accumulation follows coherent addition of non-collinear spin polarizations. Our approach was validated across multiple injector configurations, including orthogonal (45°/135°) and near collinear (15°/165°) orientations, highlighting the robustness and generality of the method. These findings establish a crucial building block for implementing spin-vector logic operations in two-dimensional materials and pave the way for scalable spintronic architectures capable of energy-based computation. Moving forward, the integration of stochastic spin vectors from low-barrier nanomagnets may enable hardware-efficient implementations of probabilistic computing models based on Heisenberg-like interactions. Future efforts may focus on extending this platform toward room-temperature operation, enhanced spin coherence control, and hybrid integration with complementary spin manipulation techniques to realize functional, reconfigurable spin-vector processors.

## Methods

### Device Fabrication

The fabrication process began with wet transferring monolayer graphene grown via chemical vapor deposition (CVD) on 4-inch $SiO_2$ wafers (purchased from Grolltex inc.) to a p++ silicon/silicon dioxide (90nm) substrate. During the wet transfer process, polystyrene was used as a carrier film to pick up the monolayer graphene films, and it was subsequently dissolved in toluene before patterning [32]. A channel width of 2 $\mu m$ was defined through e-beam lithography, followed by reactive ion etching with a $Cl_2/O_2$ mixture at 40 W for 30 seconds. Finally, ferromagnetic contacts were patterned using e-beam lithography to define contact length ranging from 200 nm to 500 nm and e-beam evaporation in two steps. The $Al_2O_3$ tunneling barrier is formed by e-beam evaporated 0.4 nm Al and naturally oxidized in air for 10 mins, followed by the Py/Ti/Au (40 nm/10 nm/30 nm) deposition in the same chamber.

### Device electrical characterization

All electrical characterization was performed in a Lakeshore FWPX Probe Station at a vacuum level of $< 10^{-5}$ Torr at room temperature, using an Agilent 4156C parameter analyzer. All devices were screened by electrical measurement of transfer characteristics before spin transport characterization.

### Device spin transport characterization

**Electrical measurements.** Spin transport measurements were performed using a lateral spin valve device fabricated on a nonmagnetic graphene channel and measured in a Quantum Design DynaCool Physical Property Measurement System (PPMS) at 20 K. The device incorporated three Permalloy (Py) ferromagnetic electrodes: two injectors ($Py_1$ and $Py_2$), oriented at distinct in-plane angles $\theta_1$ and $\theta_2$, and a third fixed detector electrode ($Py_3$). Spin currents were injected into the graphene channel from both injectors, and the resulting spin accumulation was detected using a non-local geometry. The voltage probes were placed outside the charge current path to ensure sensitivity only to spin accumulation.

**Dual source synchronized spin current injection**. To enable vectorial spin injection from the non-collinear magnetizations of $Py_1$ and $Py_2$, two synchronized sinusoidal current sources were employed. The first injector ($Py_1$) was driven using a Keithley 6221 AC current source delivering a low-noise sinusoidal current of:

$$i_1 = I_0 \sin(2\pi f t)$$

where

$$I_0 = 1 \text{ μA and } f = 17.7\ Hz$$

The second injector ($Py_2$) was driven by the voltage output of a Stanford Research Systems SR830 lock-in amplifier, connected in series with a 1 MΩ resistor to generate a matched current:

$$i_2(t) = V_{out}(t)/1\ M\Omega = I_0 \sin(2\pi f t + \varphi)$$

with $\varphi \approx 0^\circ$ due to external triggering of the lock-in amplifier by the 6221's sync output. This ensured coherent, in-phase spin injection from both $Py_1$ and $Py_2$.

**Non-local voltage detection and crosstalk suppression**. The non-local voltage $V_{NL}$(t) was measured using the same SR830 lock-in amplifier. To eliminate spurious capacitive or inductive pickup, a Stanford Research SR560 low-noise differential preamplifier was placed between the voltage probes and the lock-in input. This configuration provided high input impedance (100 MΩ) and strong common-mode rejection, reducing susceptibility to capacitive leakage, ground loops, or displacement currents. All electrical lines were verified through open-circuit and resistive standard tests prior to data acquisition to ensure signal integrity.

**Phase-shifted artifact suppression and lock-in channel validation**. To rigorously exclude phase-shifted charge-induced artifacts such as capacitive charging signals, the lock-in amplifier's quadrature (Y) channel was monitored throughout the measurement. As established by Volmer et al., non-local signals arising from capacitive displacement currents appear predominantly in the Y-channel and scale linearly with frequency:

$$V_{cc} \propto \omega\, C_{eff} V_{CM} = 2\pi f\, C_{eff} V_{CM}$$

where

$V_{CM}$ is the common-mode voltage across the detection circuit and $C_{eff}$ includes the wiring and input capacitance. In our setup, the excitation frequency was maintained at 17.7 Hz—low enough to remain below the onset of significant capacitive charging effects. No magnetic field–dependent signal or switching was observed in the Y-channel, indicating that the spin signal was confined to the in-phase (X) channel as expected for transport-limited spin accumulation. Furthermore, to minimize $C_{eff}$, cable lengths were reduced where possible and no RC low-pass filters were inserted in-line. The RC time constant of the detector circuit was estimated to be >10 ms, ensuring that phase-shifted current transients from $V_{cc}$ did not contribute within the measurement bandwidth. Back-gate voltages were applied using a Keithley 2400 Source Meter to electrostatically tune the carrier density in the graphene channel. Gate voltages ranging from –30 V to +30 V were used to modulate the spin diffusion coefficient, spin relaxation time, and spin accumulation amplitude. All measurements were repeated at multiple gate biases to confirm reproducibility and gate-tunable behavior.

**HSPICE simulation**

The spin-circuit numerical simulations are performed using HSPICE, based on a generalized $4 \times 4$ spin-circuit approach[16,28,29,33]. The spin-circuit modules are defined as sub-circuits inside HSPICE. The implementation includes non-magnet (NM) modules, which model the graphene channels, and FM|NM modules, which describe the interface between the ferromagnet (FM) and the non-magnet and are used here to represent the Permalloy (Py) ferromagnets. Both modules capture spin-dependent transport by representing each voltage node as a four-component vector: one charge and three spin components along the $(x, y, z)$ directions, and each conductance is expressed as a $4 \times 4$ matrix that incorporates coupling between charge and spin currents. Further details are provided in the Supplementary.

**Magnetic Characterization**

An ASYMFM. HM-R2 cantilever (Asylum Research, Santa Barbara, CA) was used for two pass magnetic force microscopy (MFM) measurements on an MFP-3D Bio AFM (Asylum Research, Santa Barbara, CA) system. A 5 µm × 5 µm, 128 pixel × 128 pixel scan centered on the junction of the tilted electrode was performed. A scan rate of 0.75 line per second, a scan angle of 0 deg, a nap height of 100 nm, a nap amplitude of 10 nm, a free amplitude of 20 nm, and a setpoint of 10 nm was used.

A combined Sader's method[34] /thermal method[35] calibration was performed based on the manufactured specified plain view cantilever dimensions to determine the optical lever sensitivity and cantilever stiffness. This gave a cantilever stiffness of 5 nN/nm, an optical lever sensitivity of 110 nm/V, and a fundamental resonance frequency of 82 kHz.

The cantilever tip was polarized by exposure to a strong magnetic field prior to AFM measurements. The variable field module VFM-4(Asylum Research, Santa Barbara, CA) was used to sweep the strength of the in-plane magnetic field in the direction parallel to the $Al_2O_3$ electrodes. The field was initially set to 200 G, then set to 10 G, then swept from 30 G to -200 G in steps of ~10 G. After each step in the magnetic field a 2 pass MFM topography image was captured. A FW Bell Model 5170 Gauss/Tesla meter was used to correct the magnetic field readings from the hall sensor embedded in the VFM module.

## Data availability

The datasets collected from experiments are available from the corresponding author upon reasonable request.

## Code availability

The codes used for plotting the data are available from the corresponding authors on reasonable request.

## Acknowledgements

Authors acknowledge support from the Office of Naval Research (ONR), Multidisciplinary University Research Initiative (MURI) grant N000142312708. Y.T., R. T., J.A, and Z.C. acknowledge support from the NSF Grants No. DMREF-2324203

## Author contributions

Y.T., R.T., J. A., and Z. C. designed the experiments. Y.T. and R.T. performed device fabrication, characterization, and data analysis. S.B., S.D. and K.C. assisted with theoretical analysis and HSPICE simulation. R.W. and N.D.  assisted with MFM measurements. The manuscript was written through contributions of all authors. All authors have given approval to the final version of the manuscript.

‡These authors contributed equally.

## Competing interests

The authors declare no competing interests.

**Supplementary Materials**

# Spin Vector Control for Heisenberg-Inspired Probabilistic Computing

Yuanqiu Tan ‡,[1,2] Rahul Tripathi ‡, [1,2] Saleh Bunaiyan,[4,5] Ryan Wagner,[3] Neil Dilley,[2] Kerem Camsari,[4] Joerg Appenzeller,[1,2] and Zhihong Chen*[1,2]

[1]*Electrical and Computer Engineering, Purdue University, West Lafayette, Indiana, USA*

[2]*Birck Nanotechnology Center, Purdue University, West Lafayette, Indiana, USA*

[3]*Mechanical Engineering, Purdue University, West Lafayette, Indiana, USA*

[4]*Electrical and Computer Engineering, University of California, Santa Barbara, California, USA*

[5]*Engineering Department, King Fahd University of Petroleum & Minerals (KFUPM), Dhahran 31261, Saudi Arabia*

**To whom correspondence should be addressed.*

*E-mail: zhchen@purdue.edu*

# Table of Contents

## 1. Reciprocal spin transport characteristics of non-local lateral graphene spin valves

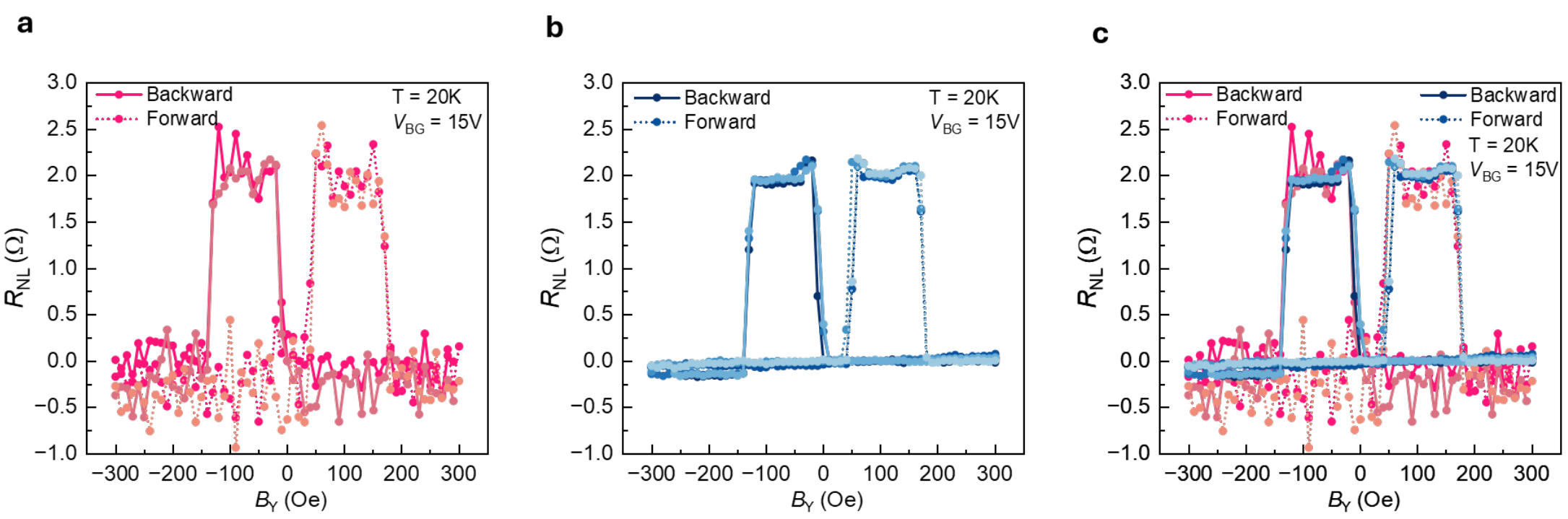

**Fig. S1. Reciprocal readout with opposite spin injection polarity. a,** Three repeated non-local spin valve measurements with spin injection through FM electrode 3 and detection at electrode 2, as schematically illustrated in Fig. 1a; **b**, Three repeated non-local spin valve measurements with spin injection through FM electrode 2 and detection at electrode 3; **c**, Combined plot of all measurements from **a** and **b**, demonstrating signal robustness, reproducibility, and Onsager reciprocal relations.

## 2. The spatial dependence of the electrochemical potential μ for spin-up and spin-down electrons in the graphene non-local spin valve configuration

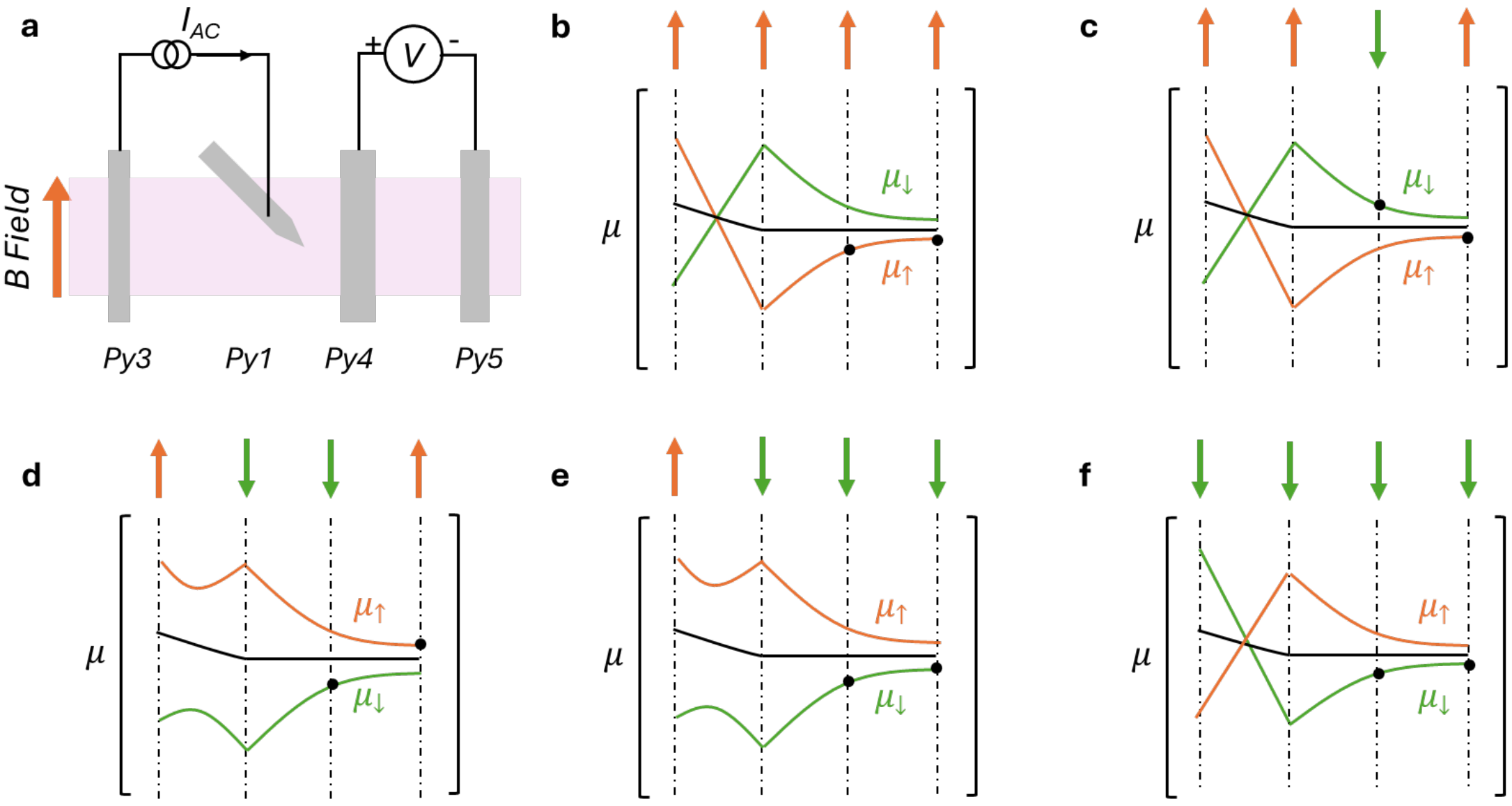

**Fig. S2. Spatial evolution of electrochemical potentials in the graphene non-local spin valve. a**, Schematic of a monolayer graphene lateral NLSV device. **b-f**, The spatial profiles of the electrochemical potentials for spin-up and spin-down electrons corresponding to sequential magnetization configurations during a magnetic-field sweep from positive to negative field under single-injection conditions. Each configuration change produces a distinct resistance step in the non-local signal. The potential difference between panels **b** and **c** corresponds to the first resistance jump labeled in Fig. 2c, while the transition from **c** to **d** corresponds to the second jump, and so forth.

For dual-injection configurations, additional resistance steps arise depending on the relative injection conditions of $I_{\mathrm{Py1}}$ and $I_{\mathrm{Py2}}$. As illustrated in Fig. 3, magnetization reversal of Py1 and Py2 introduces extra changes in the electrochemical potential profiles, resulting in an additional resistance step between panels **d** and **e** compared to the single-injection case. For example, with a relatively balanced injection from Py1 and Py2 (Fig. 3a), the electrochemical potential associated with Py1 has already transitioned to configuration **d**, while the potential associated with Py2 remains in configuration **c** due to its different switching field. This intermediate configuration produces an additional resistance step in the non-local signal. These spatial electrochemical potential traces provide a microscopic interpretation of the stepwise evolution observed in the non-local spin valve signals under different magnetic-field conditions.

## 3. Spin transport of dual spin injection between 45°/135° through non-local lateral graphene spin valve

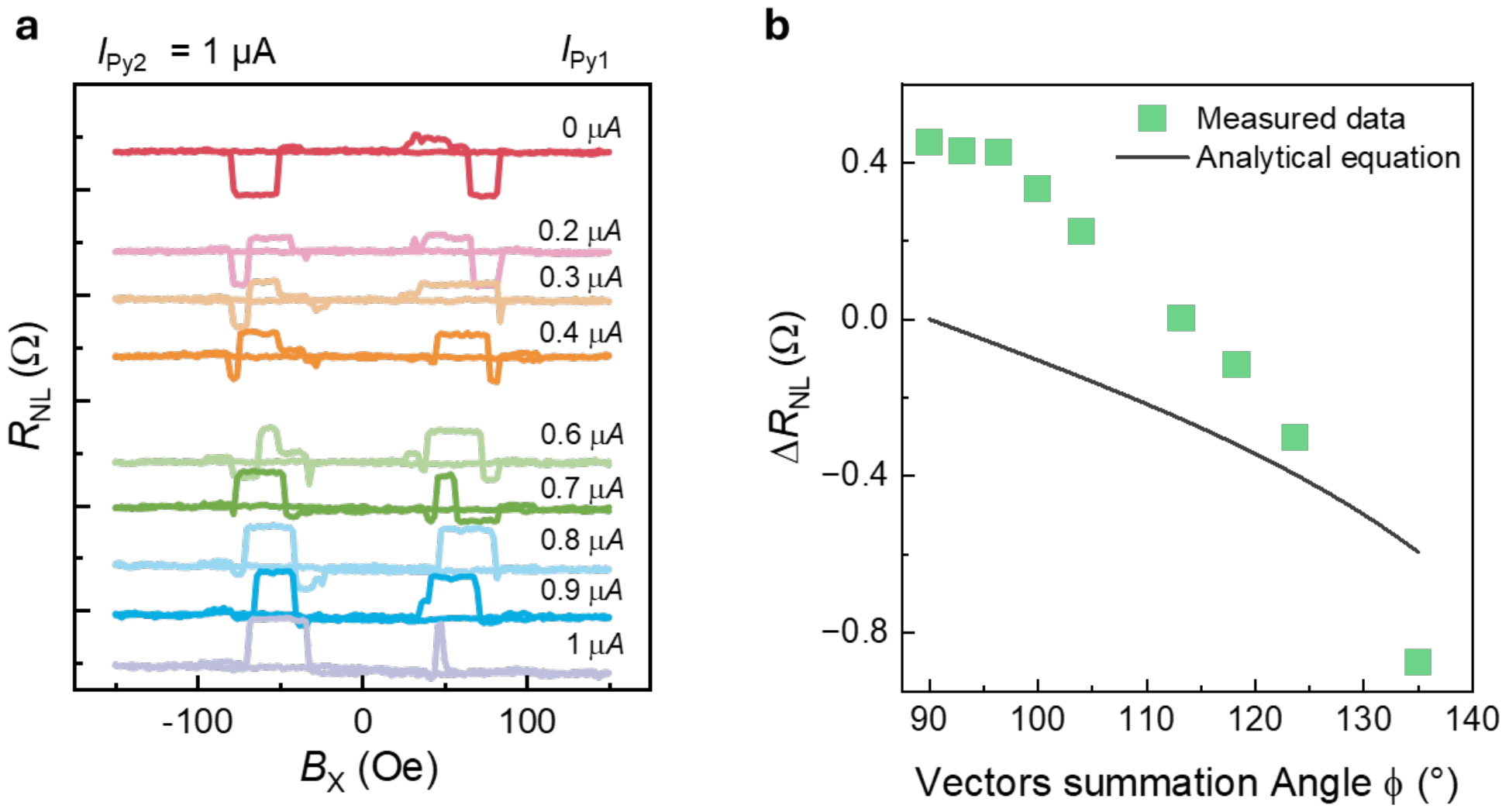

**Fig. S3. Non-colinear spin vector readout by graphene non-local spin valve. a**, Non-local spin valve signals obtained by varying $I_{Py1}$ while keeping $I_{Py2}$ = 1 μA. **b**, Corresponding $\Delta R_{NL}$ (squares) as a function of the angle ∅, with modulation achieved by changing the current injected into Py1 while holding Py2 current constant. All spin transport measurements are conducted with T = 20 K and $V_{BG}$ = 15 V.

Starting from the single injection from Py2, the resultant $\Delta R_{NL}$ presents negative values. As the current injected from Py1 is increased, the resultant spin vector rotates and $\Delta R_{NL}$ shows less negative and eventually turning positive. The non-local spin valve scan with B-field under $I_{Py1} = I_{Py2}$ = 1 μA exhibit the same switching sequence observed in Fig. 4.

## 4. HSPICE Simulation

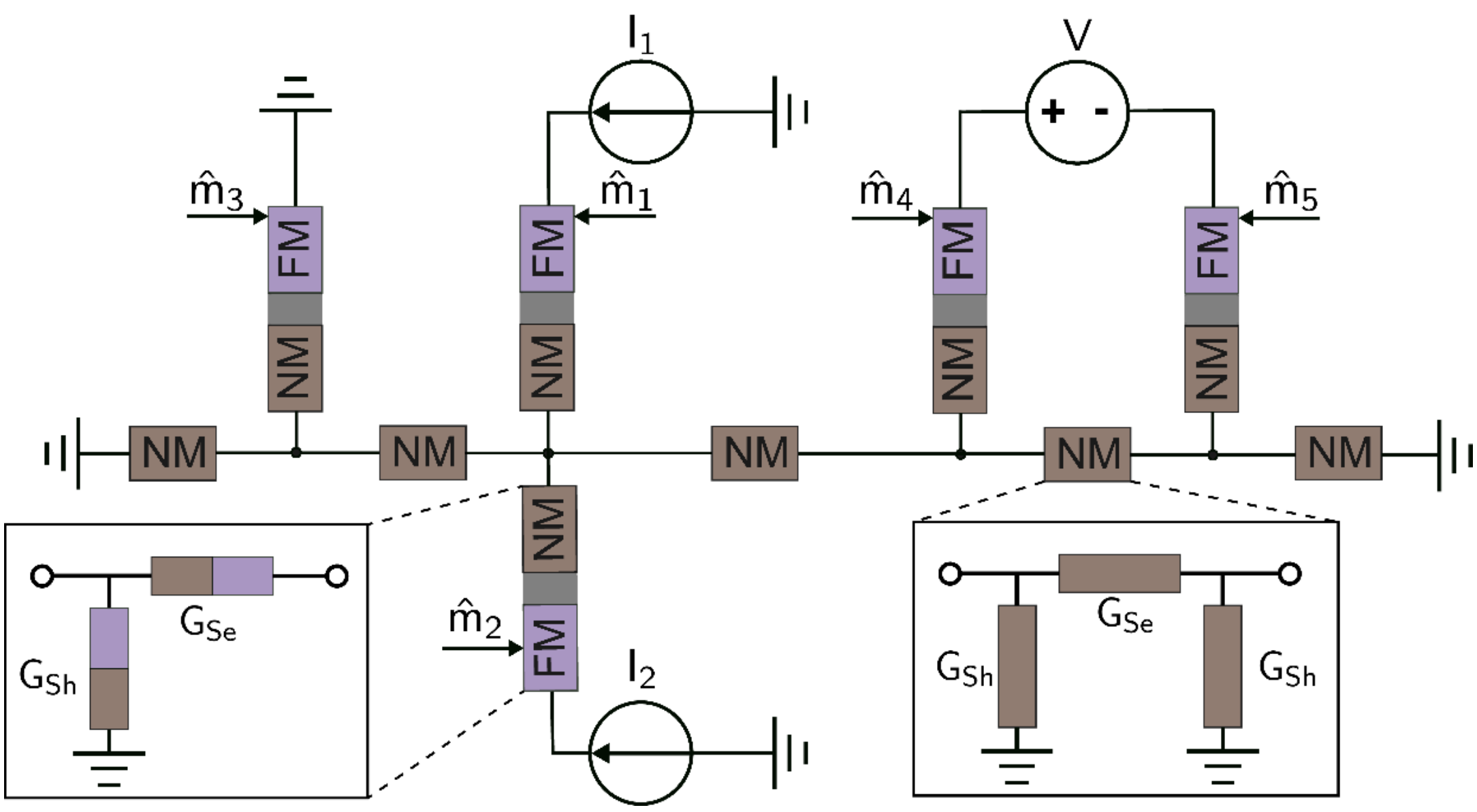

**Fig. S4. Simulated non-local spin valve using spin-circuit.** The device transport is captured by two modules: the Non-Magnet (NM) and the Ferromagnet and non-magnet interface (FM|NM). The NM module represents the graphene channels while, the FM|NM modules represent the interface between graphene and the Permalloy (Py) ferromagnet. All circuit components are described by four components: charge and spins in the $(x, y, z)$ directions. The injected spin current in the channel is controlled by $I_1$ and $I_2$, where the two injecting magnets $(\hat{m}_1, \hat{m}_2)$ are oriented along 45° and 135°, respectively. The measured non-local resistance $R_{NL}$ is described by the ratio between the non-local voltage (V) and the total injected charge current ($I_1 + I_2$).

In this section, we describe the spin-circuit approach used to emulate the non-local spin-valve device shown in Fig. 2 of the main manuscript. This modeling approach was employed to obtain both the numerical and analytical results reported in Fig. 4-5 and Fig S3.

The spin-circuit models are derived based on the work of Bauer, Brataas, and Kelly on magnetoelectric circuit theory[1]. These models were further extended into circuit formulations that can be readily integrated with standard circuit simulators (e.g., HSPICE)[2–5]. This spin-circuit approach generalizes conventional charge-based circuits to include spin-dependent transport, where each voltage node is represented by a four-component vector: one charge component and three spin components along the $(x, y, z)$ directions. Consequently, each conductance element is expressed as a $4 \times 4$ matrix.

**Non-magnet (NM):** The NM model consists of three $4 \times 4$ matrices: one series matrix $G_{Se}$ and two shunt matrices $G_{Sh}$, as shown in Fig. S4. The standard charge and spin transport are represented by $G_{Se}$, while the relaxation of spin transport is captured by $G_{Sh}$. These matrices are constructed as follows:

$$G_{Se} = \begin{bmatrix} G_c & 0 & 0 & 0 \\ 0 & G_s & 0 & 0 \\ 0 & 0 & G_s & 0 \\ 0 & 0 & 0 & G_s \end{bmatrix}, G_{Sh} = \begin{bmatrix} 0 & 0 & 0 & 0 \\ 0 & G_S' & 0 & 0 \\ 0 & 0 & G_S' & 0 \\ 0 & 0 & 0 & G_S' \end{bmatrix}$$

The charge conductance is given by $G_C = A_{NM}/(\rho_{NM}\, L_{NM})$, the series spin conductance is given by $G_s = A_{NM}/(\rho_{NM}\, \lambda_s)\,\mathrm{csch}(L_{NM}/\lambda_s)$ , and the shunt spin conductance is given as $G_S' = A_{NM}/(\rho_{NM}\, \lambda_s)\tanh(L_{NM}/2\lambda_s)$, where $A_{NM}$ is the NM's cross-sectional area, $\rho_{NM}$ is the NM's resistivity, $\lambda_s$ is the spin-diffusion length and $L$ is the NM length.

**Ferromagnet and Non-Magnet Interface (FM|NM):** The FM|NM interface model consists of two $4 \times 4$ conductance matrices: a series conductance $G_{Se}$ and a shunt conductance $G_{Sh}$. Both matrices assume that the ferromagnet's magnetization is initially aligned along the $+z$ direction. The two conductance are constructed as follows

$$G_{Se} = G_0 \begin{bmatrix} 1 & P & 0 & 0 \\ P & 1 & 0 & 0 \\ 0 & 0 & 0 & 0 \\ 0 & 0 & 0 & 0 \end{bmatrix}, G_{Sh} = G_0 \begin{bmatrix} 0 & 0 & 0 & 0 \\ 0 & 0 & 0 & 0 \\ 0 & 0 & a & b \\ 0 & 0 & -b & a \end{bmatrix}$$

where $G_0$ is the interface conductance, $P$ is the interface polarization, and $a, b$ represent the real and imaginary parts of the spin-mixing conductance, respectively. The conductance dependence on the magnetization orientation $\hat{m}_i = (\cos\phi_i \sin\theta_i\,, \sin\phi_i \sin\theta_i\,, \cos\theta_i)$ is incorporated through the rotation $G_{\{se,sh\}} = [U_R]^T\,\big[G_{\{se,sh\}}\big][U_R]$, where the rotation matrix $U_R$ is given by

$$\left[\begin{array}{c|cccc} & c & z & x & y \\ \hline c & 1 & 0 & 0 & 0 \\ z & 0 & \cos\theta & \sin\theta\cos\phi & \sin\theta\sin\phi \\ x & 0 & -\sin\theta\cos\phi & \cos\theta + \sin^2\phi(1-\cos\theta) & -\sin\phi\cos\phi(1-\cos\theta) \\ y & 0 & -\sin\theta\sin\phi & \sin\phi\cos\phi(1-\cos\theta) & \cos\theta + \cos^2\phi(1-\cos\theta) \end{array}\right]$$

**Spin-Circuit Numerical Simulation**

Using the $4 \times 4$ spin-circuit approach, we construct the non-local spin-valve setup shown in Fig. 2 of the main manuscript, which consists of five ferromagnets connected through a graphene channel. The complete circuit implementation, realized in HSPICE, is illustrated in Fig. S4. The transport in the device is captured using the NM module, which represents the graphene channels, and the FM|NM interface module, which models the interface between the graphene and the Permalloy (Py) ferromagnets.

The injected spin current in the channel is controlled by the charge-current sources $I_1$ and $I_2$, which are connected to the two injecting ferromagnets with magnetization directions $\hat{m}_1$ and $\hat{m}_2$ , oriented at 45° and 135°, respectively. The non-local resistance $R_{NL}$ is defined numerically as the ratio between the measured non-local voltage $V$ at the detector and the total injected charge current $(I_1 + I_2)$. The parameters used to generate the numerical results presented in Fig. 4 and 5 are summarized in Table S1.

Table S1 Used Parameters for spin-circuit simulation

| Parameter | Value | Unit |
|---|---|---|
| Interface polarization for detector ($P_{det}$) | 0.5 | – |
| Interface polarization for injector 1 ($P_1$) | 0.5 | – |
| Interface polarization for injector 2 ($P_2$) | 0.5 | – |
| Interface conductance ($G_0$) | 0.001 | S |
| Spin-Mixing conductance real part ($aG_0$) | 0.001 | – |
| Spin-Mixing conductance Imaginary part ($bG_0$) | 0 | – |
| NM Spin-Diffusion Length ($\lambda_s$) | 500 | nm |
| NM Resistivity ($\rho_{NM}$) | $2.35 \times 10^{-8}$ | μΩ · cm |
| NM Length ($L_{NM}$) | 500 | nm |
| NM Area ($A_{NM}$) | $5.5 \times 10^{-16}$ | $m^2$ |

**Analytical Derivation of Non-Local Resistance**

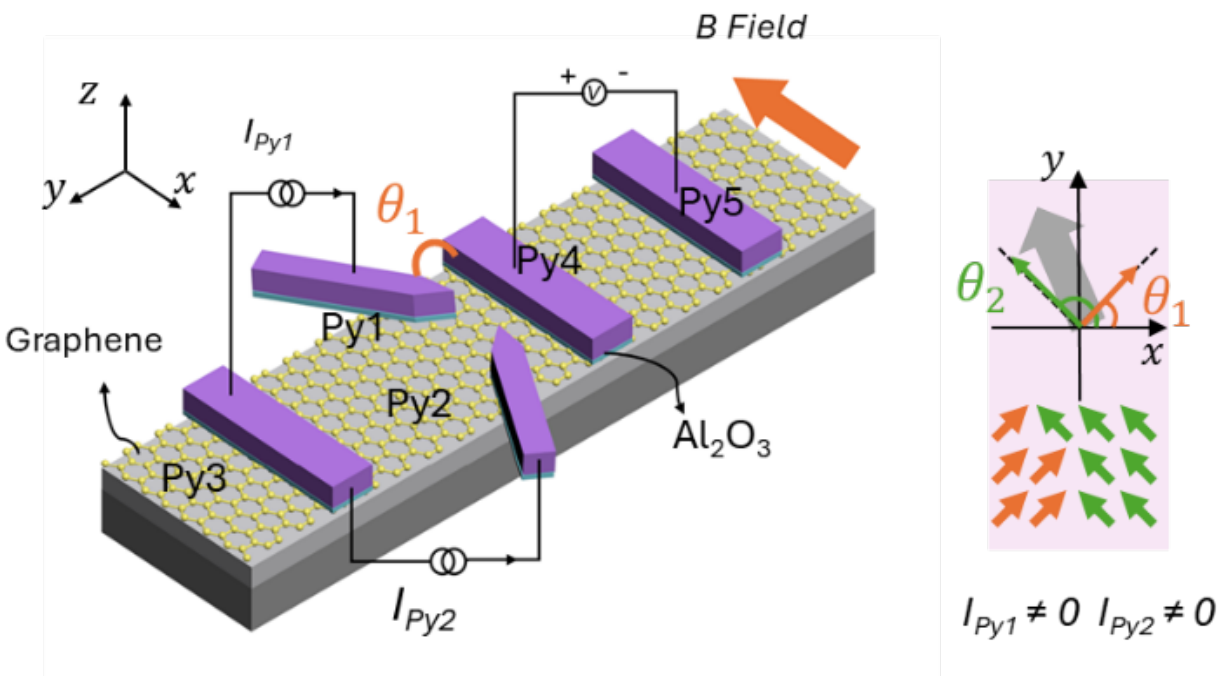

**Fig. S5. Schematic illustration of a lateral NLSV monolayer graphene device.** The dual injection electrodes are tilted to $\theta_1$ and $\theta_2$ relative to the easy-field of detector electrode in x-axis.

Building on the generalized $4 \times 4$ spin-circuit approach described above, we derive an analytical expression for the non-local resistance $R_{NL}$ using standard nodal analysis. For mathematical simplicity, we assume that each magnet has low conductance, such that the $G_0 \ll 1$ holds and the ratio between the spin-diffusion length and the channel length satisfies $\lambda_s/L_s = 1$ . Under these assumptions, the approximate analytical expression for $R_{NL}$ is given by

$$R_{NL} \approx \frac{R_{ch}\, P_{det}\, (P_1 I_1 \cos\theta_1 +\, P_2 I_2 \cos\theta_2)}{9(I_1 +\, I_2)}$$

Where $R_{ch} = (\rho_{NM}\, L_{NM})/A_{NM}$ is the graphene channel resistance, $P_{det}$ is the detector polarization, $P_{1,2}$ are the polarizations of the two injectors, and $\theta_{1,2}$ denote their orientation angles with respect to the detector. In the device shown in Fig. 2, the orientation of the first injecting magnet is fixed at $\theta_1 = 45^{\circ}$and that of the second injector at $\theta_2 = 135^{\circ}$. This analytical expression was used to obtain the results reported in Figs. 4 and 5.

## 5. Spin transport of dual spin injection between 15°/165° through non-local lateral graphene spin

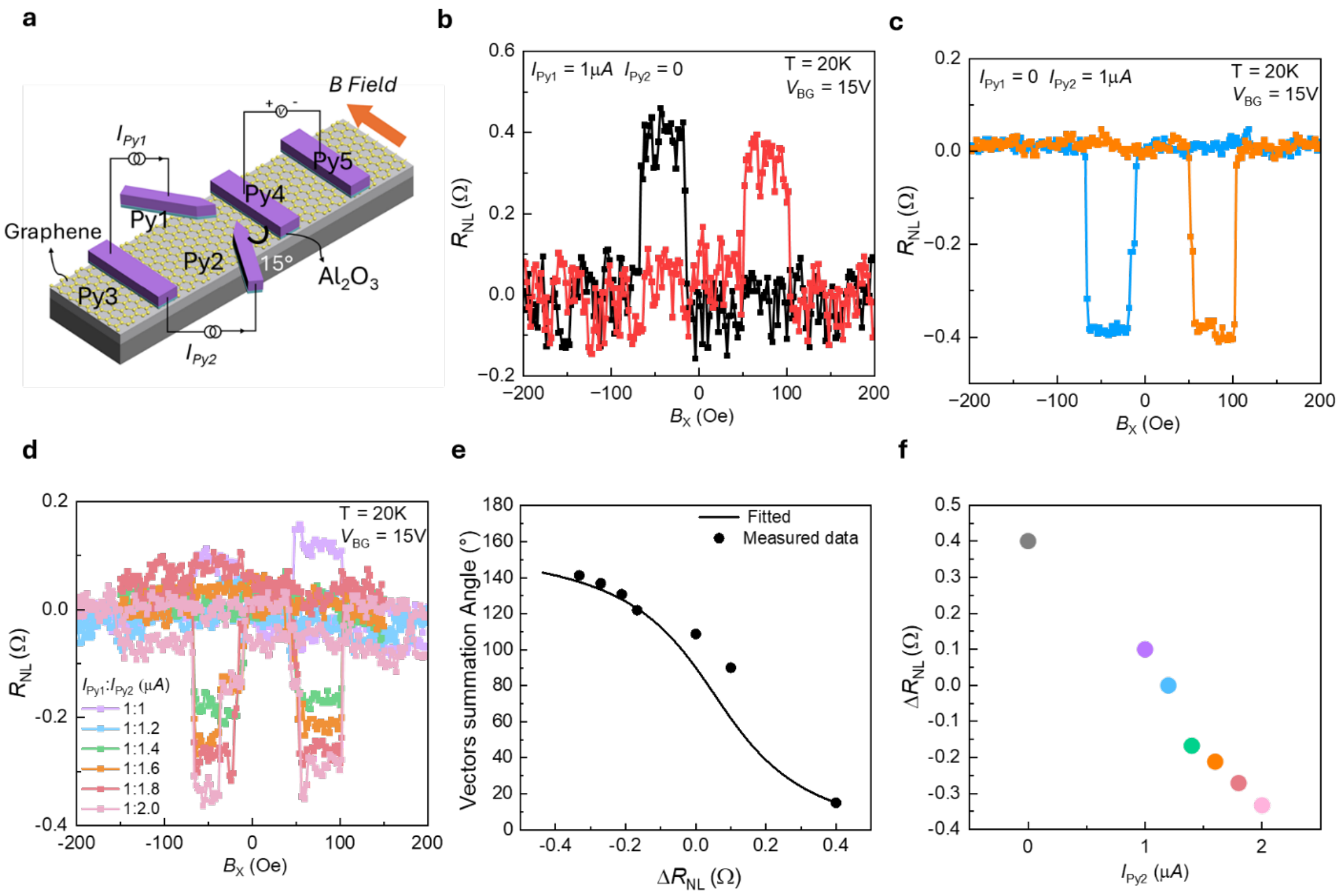

**Fig. S6. Spin vector readout by graphene non-local spin valve and theoretical modeling. a**, Schematic of lateral non-local spin valve device on monolayer graphene spin valve, with Py ferromagnetic electrodes and $Al_2O_3$ tunneling barrier; **b**, Non-local spin signal with only current injection through Py1, with T = 20 K and $V_{BG}$ = 15 V, showing nonlocal resistance of 0.4 Ω; **c**, Non-local spin signal with only current injection through Py2, showing nonlocal resistance of -0.4 Ω; **d**, Non-local spin signals with different current injections through Py2, while keeping the same current injection from Py1; **e**, Spin signal $R_{NL}$ as a function of the angle φ of the accumulated spin tuned by the dual injection. **f**, Summary of the non-local resistance with different current injections through Py2.

In this case, the ferromagnetic injector electrodes are oriented at 15° and 165° relative to the easy axis of the detector electrode. As shown in Fig. S6a, the device preserves the same graphene lateral non-local spin valve geometry as Fig. 2a. Because the injectors are positioned at 15° and 165°, the resulting spin vector exhibits a different projection onto the detector's easy axis compared to the 45° and 135° configuration, as illustrated in Fig. S6e. Nevertheless, the underlying vector summation principle remains the same: the resultant spin vector gradually transitions from a positive to a negative projection on the detector's axis as the relative injection currents vary, leading to a clear sign reversal in the non-local resistance $\Delta R_{NL}$, as shown in Fig. S6d.

Fig. S6f shows the linear relationship between $R_{NL}$ and current injected into Py2 ($I_{Py2}$), consistent with the following relations:

$$V_{NL_1} = R * I_{Py1} * \cos\Theta$$

$$V_{NL_2} = -R * I_{Py2} * \cos\Theta$$

$$V_{NL} = V_{NL_1} + V_{NL_2} = R * (I_{Py1} - I_{Py2}) \cos\Theta$$

This alternative configuration enables a distinct mode of spin vector summation and further demonstrates the robustness and tunability of our dual-injection method for manipulating spin orientations over a broader vector space.